\documentclass[12pt,preprint]{aastex}

\shorttitle{Global Features of The X-Ray Light Curves of {\it SWIFT} GRBs}
\shortauthors{Shao et al.}

\begin{document}

\title{UNDERLYING GLOBAL FEATURES OF THE X-RAY LIGHT CURVES OF {\it SWIFT} GAMMA-RAY BURSTS}

\author{Lang Shao, Yi-Zhong Fan, and Da-Ming Wei}
\affil{Purple Mountain Observatory, Chinese Academy of Sciences, Nanjing 210008, China}
\email{lang@pmo.ac.cn}

\begin{abstract}
With its rapid response, {\it Swift} has revealed plenty of unexpected properties of gamma-ray bursts (GRBs). With an abundance of observations, our current understanding is only limited by complexity of early X-ray light curves. In this work, based on the public {\it Swift} data of 150 well-monitored GRBs with measured redshifts, we find some interesting global features in the rest-frame X-ray light curves. The distinct spectral behaviors between the prompt emission and the afterglow emission implies dissimilar radiation scenarios. Interestingly, an unforeseen plateau is exhibited in the prompt X-ray light curves despite the presence of complex spikes, which might indicate the presence of a steady central engine. In particular, the seemingly continuous evolution with a single power law from the prompt to the afterglow of most GRBs might place strong constraints on the theoretical models.

\end{abstract}


\section{Introduction}

Ever since gamma-ray bursts (GRBs) were first reported as flashes of cosmic gamma rays lasting for less than a tenth of a second to more than tens of seconds \citep{klebesadel73}, they have shown a great diversity. Unfortunately, their energy source and radiation mechanism remain highly speculative \citep[e.g., see][for a review]{fishman95}. The first detection of X-ray \citep{costa97} and optical \citep{vanparadijs97} afterglows that faded as a single power law $\propto t^{-1.3}$, strongly favored the prior relativistic fireball models \citep{wijers97}. Even though the origin of prompt gamma-rays remained uncertain, the longer-wavelength afterglows, especially the optical afterglows, have been found broadly consistent with fireball models \citep[e.g., see][for a review]{meszaros02}.

The fundamental understanding of GRB afterglows has been questioned by the {\it Swift} mission \citep{gehrels04}, which has revealed several unexpected behaviors in the X-ray afterglows \citep[e.g., see][for a review]{zhang07}. Some bursts have an initial steep decline, followed by a shallow decline a few hundred seconds later \citep[e.g.,][]{tagliaferri05}. Others have erratic flares with strong spectral variations \citep[e.g.,][]{burrows05,chincarini07}. As a group, they were proposed to have some canonical behaviors \citep{zhang06,nousek06}. Contradictorily, some other bursts also show evidences for a single power-law decline \citep{liang09}, which are consistent with the pre-{\it Swift} observations \citep{costa99}.

With the X-ray afterglows being puzzlingly diverse, it is a great challenge to produce an applicable and self-consistent physical understanding \citep[e.g., see][for a review]{zhang07}. Previous works on GRB diversity were based on relatively restricted samples, e.g., 27 bursts in \citet{nousek06},  with only 10 measured redshifts; 33 bursts in \citet{chincarini07}, with only 9 measured redshifts; and 19 bursts in \citet{liang09}, with only 12 measured redshifts. With more than 500 {\it Swift} GRBs detected up to now, we realize that it is urgent to revisit the diversity issue with an extended sample.

Using a collection of 150 {\it Swift} GRBs with well detected X-ray afterglows and known redshifts, we find that there exist some underlying global features in the rest-frame X-ray light curves, which might clarify the diversity issue and provide some strong constraints on the theoretical models. The structure of this Letter is as follows: the sample and data analysis are presented in Section~\ref{sec:sample}, the rest frame light curves are interpreted in Section~\ref{sec:ugf}, and the discussions and conclusion are given in Section~\ref{sec:discussion}.

\section{Sample and Methodology} \label{sec:sample}

{\it Swift} is a multiwavelength observatory \citep{gehrels04}, which carries three instruments: Burst Alert Telescope (BAT), X-Ray Telescope (XRT), and Ultraviolet/Optical Telescope (UVOT). We made extensive use of the automated BAT-XRT products provided by the UK {\it Swift} Science Data Centre \citep{evans10}, the online BAT GRB Event Data Processing Report provided by Taka Sakamoto and Scott D. Barthelmy, and the online big table of all well-localized GRBs provided by Jochen Greiner\footnote{http://www.mpe.mpg.de/$\sim$jcg/grbgen.html, which is mainly our source for measured redshifts.}.

\begin{figure}
\begin{center}
\includegraphics[scale=0.8]{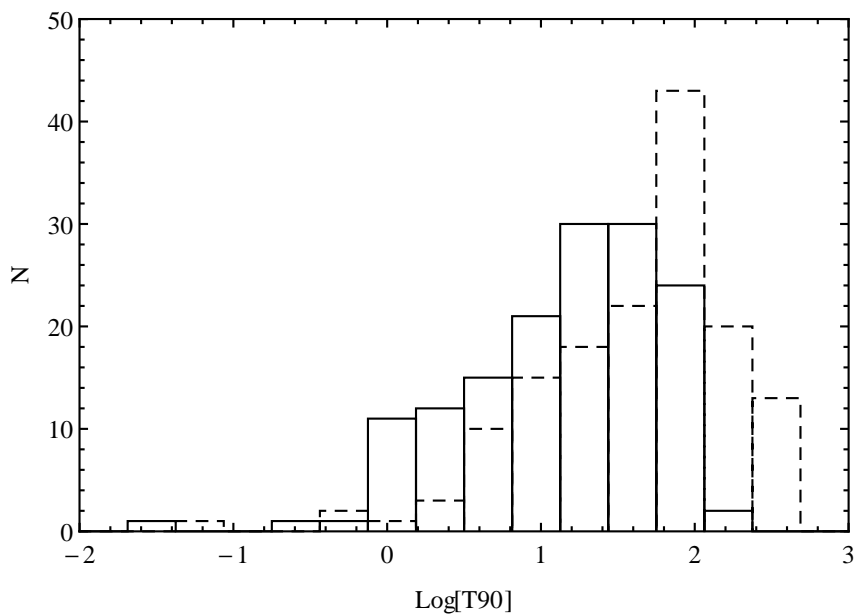}
\end{center}
\caption{Distribution of the BAT burst durations $T_{90}$ measured in observer's (dashed histogram) and rest (solid histogram) frame, respectively.
\label{fig:t90}}
\end{figure}

In Table \ref{tab:sample}, we list the 150 bursts from GRB 050126 to GRB 100425A with measured redshifts. We find the mean redshift $<z>\,\sim$ 2.14 (with the median $\sim 1.93$), which is significantly smaller than the previously evaluated (e.g., $<z>\,\sim$ 2.8 for 16 bursts by \citealp{jakobsson06}). The distributions of the BAT duration $T_{90}$ both in the observer and rest frame are shown in Figure \ref{fig:t90}. The mean value of $T_{90}$ in observer's frame is 80.4 s, consistent with that deduced in a sample of 237 bursts by \citet{sakamoto08}. With the 150 measured redshifts in our sample, we find that the mean value of $T_{90}$ in the rest frame is 29.2 s. Shorter bursts increase significantly in number, but no positive classification can be made. Even so, as indicated in Figure \ref{fig:t90}, there is an apparent oddball, GRB050509B, which has an extremely short duration $T_{90}\sim 0.04$ s in the rest frame.

To view the overall behaviors of GRBs in their rest frame, we take advantage of the {\it Swift} Burst Analyser, which provides the combined BAT-XRT light curves in the form of flux density $F_\nu(\nu)$ at 10 keV in the observer's rest frame \citep{evans10}. With the measured redshift $z$,  the isotropic spectral luminosity at a given photon frequency $\nu_0$ (which is 10 keV in this work) in the rest frame can be calculated by
\begin{eqnarray}
L_\nu(\nu_0,t)&=&L_\nu(\nu_e,t)\left(\nu_0\over \nu_e\right)^{1-\Gamma(t)}\nonumber \\  &=& 4\pi F_\nu(\nu_0,t_0) D_{\rm L}^2(z)(1+z)^{\Gamma(t_0)-2},
\end{eqnarray}
where $\nu_e=\nu_0(1+z)$ is the emitted photon energy in the rest frame for the observed photon energy $\nu_0$ and $t=t_0/(1+z)$ is the time measured in the rest frame related to the time $t_0$ measured in the observer's frame. Here, a single power-law spectrum has been assumed $F_\nu(\nu,t)\propto\nu^{1-\Gamma(t)}$, where $\Gamma(t)$ is the time-dependent photon index, which is available by the {\it Swift} Burst Analyser \citep{evans10}. Throughout this work, the luminosity distance $D_{\rm L}(z)$ is calculated by assuming the cosmological parameters $H_0=71\,{\rm km}\, {\rm s}^{-1}\, {\rm Mpc}^{-1}$, $\Omega_M=0.27$, and $\Omega_\Lambda=0.73$. According to the error propagation rule, the uncertainty in the isotropic spectral luminosity can be given by
\begin{equation}
\sigma_{L_\nu}^2=\left[4\pi D_{\rm L}^2 (1+z)^{\Gamma-2}\right]^2\left[\sigma_{F_\nu}^2+F_\nu^2\ln^2(1+z)\sigma_\Gamma^2\right],
\end{equation}
where $\sigma_{F_\nu}$ and $\sigma_\Gamma$ are the uncertainties in the measured flux density and photon index, respectively, and the uncertainty in GRB redshift measurements is considered negligible.

\section{Underlying Global Features} \label{sec:ugf}

\begin{figure}
\begin{center}
\includegraphics[scale=0.6]{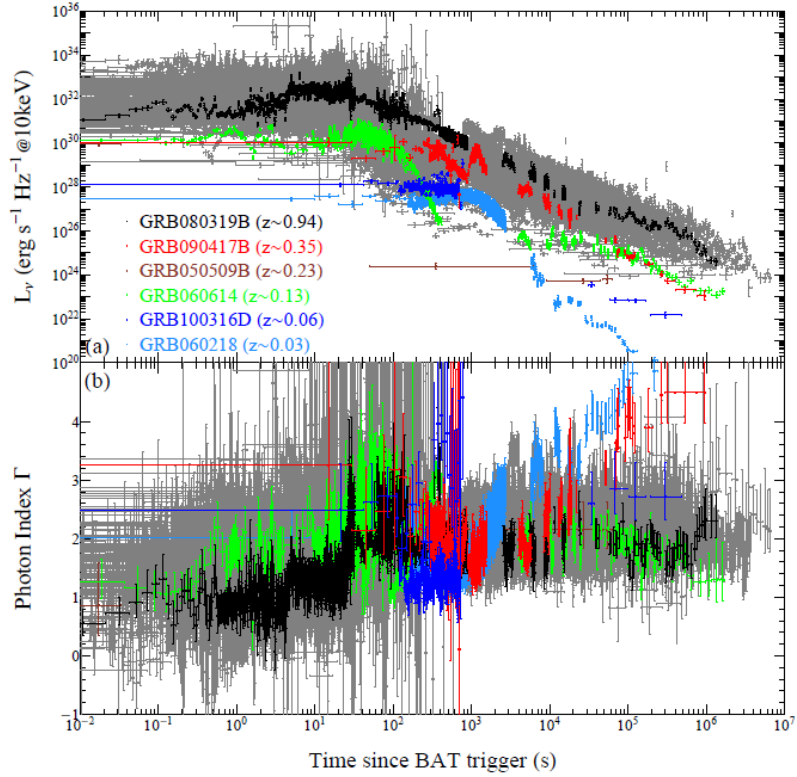}
\end{center}
\caption{X-ray light curves (top panel) and spectral indices (bottom panel) of 150 {\it Swift} GRBs in the rest frame. Six individual bursts with different redshifts are indicated with different colors from top to bottom. The rest are plotted in gray color \citep{caprio05}.  \label{fig:restframelc}}
\end{figure}

In Figure \ref{fig:restframelc}, the calculated X-ray light curves from 0.01 s to $10^7$ s after BAT trigger at 10 keV in the rest frame with the spectral evolution of 150 {\it Swift} GRBs are plotted together. Interestingly, some underlying global features are revealed in both the light curves and spectral evolution, despite the fact that some bursts may have some non-trivial behaviors. The bulk of the light curves build up a plateau lasting for about tens of seconds with significantly spectral softening and a following power-law decline with almost invariant spectra until the detection limit is reached. This feature, for instance, is well exhibited by the remarkable naked-eye burst GRB080319B\footnote{The previously reported discontinuities between the BAT-XRT combined light curves have been minimized by accounting for the spectral evolution. For details of how these light curves were produced, see \citet{evans10}.} \citep{racusin08}. There are also several apparent outliers, e.g., GRB060218, GRB100316D, and GRB060614. The former two showed clear evidences of an associated supernova \citep[e.g.,][]{soderberg06,chornock10,starling10,fan10}, while on the contrary the latter showed unexpected evidences of no associated supernova \citep[e.g.,][]{fynbo06,dellavalle06}. Another two peculiar ones are the extremely short burst GRB050509B and the optically dark burst GRB090417B. The former had an overall much lower luminosity, and the latter showed evidences of dust extinction and scattering, revealed by the strong spectral softening \citep{holland10}, as predicted by \citet{shao07} and \citet{shao08}. Interestingly, those outliers tend to have a lower redshift.

To explore the underlying global features mentioned above, we fit a broken power-law model to the light curves with the 145 GRBs (i.e., without the apparent outliers, GRB050509B, GRB060218, GRB060614, GRB090417B, and GRB100316D), assuming a four-parameter function
\begin{equation}
\label{eq:lc1}
L_\nu(t)=\cases{ L_{\nu,{\rm b}}\left({t \over t_{L,{\rm b}}}\right)^{\alpha_1}, & $t<t_{L,{\rm b}}$, \cr
L_{\nu,{\rm b}}\left({t \over t_{L,{\rm b}}}\right)^{\alpha_2}, & $t\geq t_{L,{\rm b}}$, \cr}
\end{equation}
where ($t_{L,{\rm b}}$,$L_{\nu,{\rm b}}$) is the transition position between the two power laws with the indices as $\alpha_1$ and $\alpha_2$, respectively. The best-fitting results (top panel in Figure \ref{fig:restframelc2}, dashed line) are $t_{L,{\rm b}}=15.8\pm0.4$ s, $L_{\nu,{\rm b}}=(1.2\pm0.1)\times10^{32}\,{\rm erg}\,{\rm s}^{-1}{\rm Hz}^{-1}$, $\alpha_1=-0.04\pm0.01$, and $\alpha_2=-1.47\pm0.01$. The goodness of fit is measured by the adjusted coefficient of determination, $\bar{R}^2\sim0.99928$, which is very close to 1.0 and indicates a good fit. Separately, we fit the spectral evolution with another four-parameter function
\begin{equation}
\label{eq:phontofit1}
\Gamma(t)=\cases{ \Gamma_b+\delta_1\log\left({t\over t_{\Gamma,{\rm b}}}\right), & $t<t_{\Gamma,{\rm b}}$, \cr
\Gamma_b+\delta_2\log\left({t\over t_{\Gamma,{\rm b}}}\right), & $t\geq t_{\Gamma,{\rm b}}$, \cr}
\end{equation}
where ($t_{\Gamma,{\rm b}}$,$\Gamma_{\rm b}$) is the transition position between two linear segments in the log-linear plot, with the slopes as $\delta_1$ and $\delta_2$, respectively. The best fitting results (bottom panel in Figure \ref{fig:restframelc2}, dashed line) are $t_{\Gamma,{\rm b}}=131.2\pm4.2$ s, $\Gamma_{\rm b}=2.10\pm0.01$, $\delta_1=0.47\pm0.01$, and $\delta_2=-0.03\pm0.01$ ($\bar{R}^2\sim0.93809$).

At this point, based on the light curves and spectral evolution history, two emission ingredients can be distinguished in the X-ray emission: an early plateau with significant spectral softening and a following power-law decline without spectral variance. Given that $t_{L,{\rm b}}\sim<T_{90}>$, this strongly suggests that the former is connected with the prompt emission and the latter is connected with the afterglow emission. Since these two components have been thought to arise from different radiation regions in the relativistic ejecta, the seemingly nice transition in the light curves may raise a fine-tuning issue. Nevertheless, the discrepancy between $t_{L,{\rm b}}$ and $t_{\Gamma,{\rm b}}$ implies that the interpretation is more complex.

To further explore the details at the junction, we now introduce two additional breaks both in the light curve and in the spectral evolution. Therefore, we have an eight-parameter model for the light curve defined by
\begin{equation}
\label{eq:lcfit2}
L_\nu(t)=\cases{
L_1\left({t \over t_1}\right)^{\alpha_1}, & $t<t_1$, \cr
L_1\left({t \over t_1}\right)^{\alpha_2}, & $t_1\leq t < t_2$, \cr
L_3\left({t \over t_3}\right)^{\alpha_3}, & $t_2\leq t < t_3$, \cr
L_3\left({t \over t_3}\right)^{\alpha_4}, & $t \geq t_3$, \cr
}
\end{equation}
where, ($t_1$,$L_1$) and ($t_3$,$L_3$) are the positions of the first and third break points, respectively, and $\alpha_1$, $\alpha_2$, $\alpha_3$, and $\alpha_4$ are the sequential four power-law indices. Given these eight free parameters, the position of the second break point is fixed according to Equation (\ref{eq:lcfit2}) and can be determined by
\begin{eqnarray}
  t_2 &=& \left({L_1 \over L_3}{t_3^{\alpha_3} \over {t_1^{\alpha_2}}} \right)^{1\over {\alpha_3-\alpha_2}}, \label{eq:lct2} \\
  L_2 &=& \left[ { L_1^{\alpha_3} \over {L_3^{\alpha_2}}}\left( {t_3 \over t_1}\right)^{\alpha_2\alpha_3}  \right]^{1\over {\alpha_3-\alpha_2}}. \label{eq:lcl2}
\end{eqnarray}
The best-fitting results (top panel in Figure \ref{fig:restframelc2}, thick line) are $t_1=26.5\pm0.4$ s, $L_1=(1.4\pm0.1)\times10^{32}\,{\rm erg}\,{\rm s}^{-1}{\rm Hz}^{-1}$, $t_3=620.3\pm80.9$ s, $L_3=(4.3\pm0.7)\times10^{29}\,{\rm erg}\,{\rm s}^{-1}{\rm Hz}^{-1}$, $\alpha_1=0.02\pm0.01$, $\alpha_2=-2.58\pm0.03$, $\alpha_3=-0.90\pm0.07$, and $\alpha_4=-1.35\pm0.01$ ($\bar{R}^2\sim0.99932$). Given the best-fitting parameters, we have $t_2\sim154.6$ s, and $L_2\sim 1.5\times10^{30}\,{\rm erg}\,{\rm s}^{-1}{\rm Hz}^{-1}$ according to Equations (\ref{eq:lct2}) and (\ref{eq:lcl2}).

Similarly, we have another eight-parameter model for the spectral evolution defined by
\begin{equation}
\label{eq:phontofit2}
\Gamma(t)=\cases{
\Gamma_1+\delta_1\log\left({t \over t_{1}}\right), & $t<t_1$, \cr
\Gamma_1+\delta_2\log\left({t \over t_{1}}\right), & $t_1\leq t < t_2$, \cr
\Gamma_3+\delta_3\log\left({t \over t_{3}}\right), & $t_2\leq t < t_3$, \cr
\Gamma_3+\delta_4\log\left({t \over t_{3}}\right), & $t \geq t_3$, \cr
}
\end{equation}
where ($t_1$,$\Gamma_1$) and ($t_3$,$\Gamma_3$) are the first and third break points, respectively, between linear segments in the log-linear plot, and $\delta_1$, $\delta_2$, $\delta_3$, and $\delta_4$ are the sequential four slopes. Given these eight free parameters, the position of the second break point is also fixed according to Equation (\ref{eq:phontofit2}) and can be determined by
\begin{eqnarray}
  \log(t_2) &=& {{\Gamma_3-\Gamma_1}\over{\delta_2-\delta_3}}+{1\over {\delta_2-\delta_3}}\log\left({{t_1^{\delta_2}}\over{t_3^{\delta_3}}} \right), \label{eq:spet2}\\
  \Gamma_2 &=& {{\Gamma_1\delta_3-\Gamma_3\delta_2} \over {\delta_3-\delta_2}}+{{\delta_2\delta_3}\over{\delta_3-\delta_2}}\log\left( {t_3 \over t_1}\right). \label{eq:spegamma2}
\end{eqnarray}
The best-fitting results (bottom panel in Figure \ref{fig:restframelc2}, thick line) are $t_1=13.4\pm0.7$ s, $\Gamma_1=1.45\pm0.01$, $t_3=421.8\pm29.2$ s, $\Gamma_3=1.99\pm0.01$, $\delta_1=0.17\pm0.01$, $\delta_2=0.79\pm0.01$, $\delta_3=-0.47\pm0.05$, and $\delta_4=0.04\pm0.01$ ($\bar{R}^2\sim0.93993$). Given the best-fitting parameters, we have $t_2\sim131.0$ s and $\Gamma_2\sim 2.23$ according to Equations (\ref{eq:spet2}) and (\ref{eq:spegamma2}).

The goodness of fit has been improved for both the light curve and the spectral evolution, though the spectral evolution still appears to be a little more deviant than is presumed by the fitting model. The general agreement of the three break times of the light curve and those of the spectral evolution, in spite of the fact that the two fittings are made separately, justifies the existence of the extra breaks we have just assumed. Therefore, the four-segment fitting model (with eight free parameters) is a more realistic interpretation for the underlying global feature, which, hereby, is composed of an early plateau ($\propto t^{0}$) with a mild softening ($\Delta \Gamma \sim 0.5$ on average), a following steep decline ($\propto t^{-2.6}$) with a further severe softening ($\Delta \Gamma \sim 1.0$ on average), a newly emerging shallow decline ($\propto t^{-0.9}$) with a slight spectral hardening ($\Delta \Gamma < 0.5$ on average), and a following single power-law decline ($\propto t^{-1.4}$) without spectral variation ($\Gamma\sim 2$) until the detection limit is reached. However, as indicated by Figure \ref{fig:restframelc2}, the four-segment model is not overwhelmingly better than the two-segment model, since the steep decline and the shallow decline of most bursts are interlacing with each other and cannot be well defined in both the light curves and spectral evolution. For some bursts (e.g., GRB080319B), the existence of the shallow decline and the steep decline is debatable.

\section{Discussion and Conclusion} \label{sec:discussion}

GRBs have been exposed to be diverse both in their observed prompt emission and afterglow emission. As a result, the physical interpretation has remained controversial. In this work, based on the X-ray data \citep{evans10} of 150 well-monitored {\it Swift} GRBs with measured redshifts, we find, in the rest frame, some underlying global features concealed in the observed complex X-ray light curves and the spectral evolution, which could clarify the diversity issue.

\begin{figure}
\begin{center}
\includegraphics[scale=0.6]{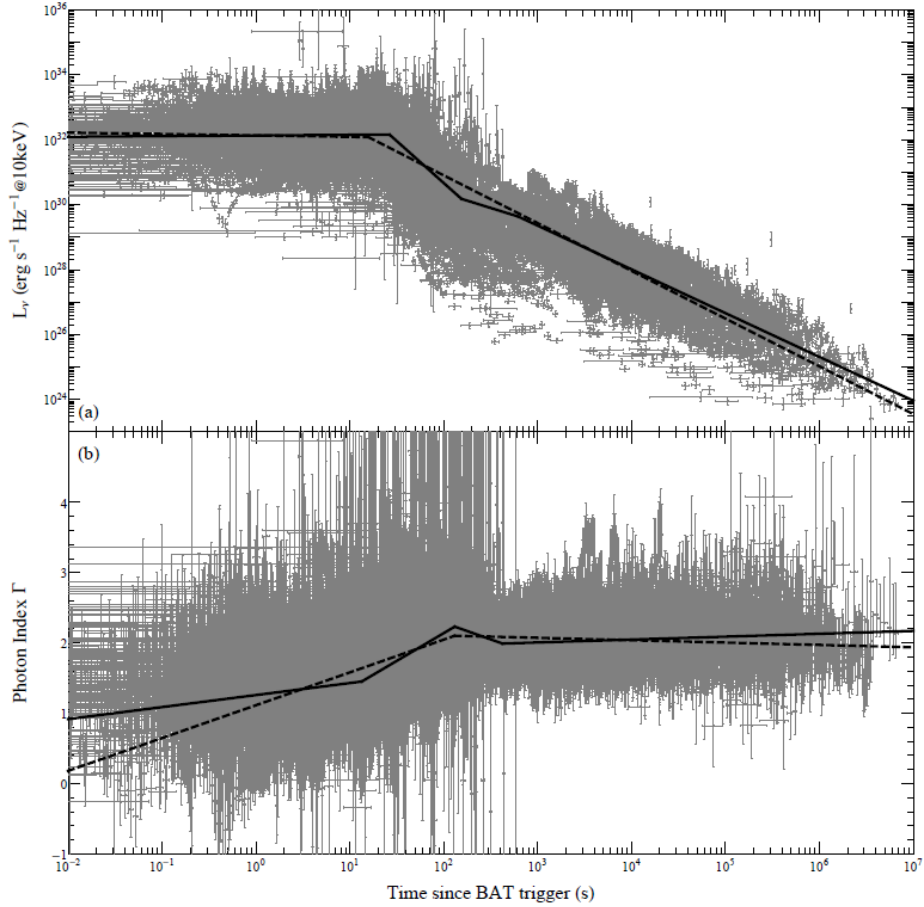}
\end{center}
\caption{X-ray light curves (top panel) and spectral indices (bottom panel) of 145 {\it Swift} GRBs (described in the text) in the rest frame. The best-fitting models with four parameters and eight parameters are shown with dashed lines and solid lines, respectively.
\label{fig:restframelc2}}
\end{figure}

Generally speaking, two distinct and consequent emission ingredients unequivocally exist in the X-ray light curves as explicitly revealed by the spectral differences shown in Figure \ref{fig:restframelc2}, which can be simply recognized as the prompt emission with apparent spectral softening and the afterglow emission without apparent spectral evolution, respectively. The most straightforward global features are the early plateau ($\propto t^{\alpha_1}$ with $\alpha_1\sim 0$ on average) with a mild spectral softening (with a photon index increase $\Delta\Gamma\sim 0.5$ on average) and the late single power-law decline ($\propto t^{\alpha_4}$ with $\alpha_4 \sim -1.4$ on average) without spectral variation (with a constant photon index $\Gamma\sim 2$ on average) up to $t\sim 10^{7}$ s, which are the previously known prompt emission and afterglow emission, respectively.

The transition from the prompt emission to the afterglow emission undergoes two distinct phases: a steep decline ($\propto t^{\alpha_2}$ with $\alpha_2 \sim -2.6$ on average) with a severe spectral softening ($\Delta\Gamma\sim 1.0$ on average) and a shallow decline ($\propto t^{\alpha_3}$ with $\alpha_3 \sim -0.9$ on average) with a mild spectral hardening to agree with the invariant spectra in the afterglow emission. The temporal feature is consistent with the canonical behavior previously proposed \citep{zhang06,nousek06}. The spectral feature supports the speculation that the steep decline is connected with the prompt emission, while the shallow decline is more relevant to the afterglow emission. On the other hand, the erratic X-ray flares reportedly existing in half of the bursts are not self-explanatory in our analysis. Since they exhibit significant spectral softening and their peak luminosity follows a steep power-law decline $L_{\rm pk}\propto t^{-2.7}$ statistically \citep{chincarini10}, they should be connected with the steep declines, which also explains their analogy to the prompt emission.

Furthermore, as proposed earlier the spectral feature could be a crucial signature for discriminating the prompt emission from the afterglow emission. For this reason, the high-energy emission detected by the {\it Fermi} satellite which normally decays as a single power law with an index $\alpha \sim -1.4$ and shows no significant spectral evolution could be more convincingly interpreted as the emission of the external forward shock \citep[e.g.,][]{zou09,kumar09,gao09,ghisellini10}.

Interestingly, despite being composed of complex spikes, the superposition of the prompt X-ray light curves of different bursts exhibits an {\it unexpected} plateau lasting about $t_1\sim 30$ s on average. This might signal a rather steady central engines. Furthermore, it indicates that the variabilities displayed by the complex spikes may just manifest the internal instabilities in the relativistic outflows launched by the central engines. For the family of black-hole central engines, this may shed some lights on the process of accreting onto a nascent black hole. We assume that the accretion process does not significantly alter the mass of the black hole ($M_{\rm bh}$) and the accretion timescale is that of the free fall $t\sim R^{3/2}\sqrt{G M_{\rm bh}}\sim 50\,{\rm s}\,R_{10}^{3/2}(M_{\rm bh}/2 M_\sun)^{-1/2}$, where $G$ is the Newton's constant and $R$ is the initial distance of the accreting matter to the central engine. Then we simply estimate the accretion rate as ${\rm d}M/{\rm d}t\propto n R^2 {\rm d}R/{\rm d}t\propto n R^{3/2}$, where $n$ is the number density of the star material being accreted. The ``steady'' prompt emission may suggest that ${\rm d}M/{\rm d}t \propto t^0$, which could in turn imply that $n\propto R^{-3/2}$. Such a density profile should hold up to a radius $R\sim 5\times 10^9\,{\rm cm}\,(M_{\rm bh}/2 M_\sun)^{1/2}$, beyond which the density profile should be steepened to $n\propto R^{-5.4}$ to account for the steep decline $\propto t^{-2.6}$, supposing the luminosity of the X-ray emission is still proportional to\footnote{In this scenario, the steep X-ray decline is attributed to the weakening of the activity around the central engine, as proposed by \citet{fan05}. In the high-latitude emission (i.e., curvature effect) model \citep[see][and references therein]{zhangbb07}, the decline should be $(t-t_0)^{-(1+\Gamma)}$, where $t_0$ is the ejection time of the last main pulse in which high latitude emission dominates over that of all other pulses. For the global feature of the steep decline  found in Figure \ref{fig:restframelc2} this model works for $t_0 \sim 0$.} ${\rm d}M/{\rm d}t$. The steep decline continues up to $t_2\sim 155$ s corresponding to $R\sim 2\times 10^{10}$ cm. On the other hand, for a magnetar-like central engine, the plateau phase can be naturally accounted for if the prompt emission is powered by the dipole radiation of the magnetar $L_{\rm dip}\propto (1+t/\tau_0)^{-2}\propto t^0$ for $t\ll \tau_0$, where $\tau_0$ is the so-called spin-down timescale \citep[e.g.,][]{pacini67,gunn69}. The quick decline phase could indicate that either the magnetar has collapsed due to significant accretion or the efficiency of converting the wind energy of the spin-down ($t>\tau_0$) magnetar decreases with time.

Last but not least, we need to mention that the seemingly continuous evolution with a single power law from the prompt to the afterglow phase of most GRBs, which was first hinted by the {\it BeppoSAX} satellite \citep{costa99}, might provide some strong constraints on the theoretical models. Otherwise, a severe fine-tuning would be needed. Meanwhile, the well-known canonical behavior seems to be overemphasized. The presence of a steep decline followed by a shallow decline ($\propto t^{\alpha}$ with $\alpha \geq -0.5$) in the early X-ray light curves might not be as prevailing as previously thought. Even though, a statistical bias may exist with respect to the bursts with a well-determined redshift, which should be brighter and less affected by dust extinction \citep[e.g.,][]{fynbo10}. Interestingly, the bursts in our sample that exhibit a single power-law decline tend to have a relatively larger luminosity. While the bursts that have puzzlingly different features, especially those that clearly exhibit a steep decline followed by a shallow decay, tend to have a much lower luminosity during the shallow decline, which may give some hint on the origin of the shallow declines and disfavor the energy injection models.

\acknowledgments

We are very grateful to the referee for insightful comments and Z.-G. Dai, B. Zhang, E.-W. Liang, S. Covino, and Z.-P. Jin for helpful communications. L.S. is grateful to N. Mirabal for proofreading the manuscript and M. Caprio for helping on the usage of LevelScheme package. This work made use of data supplied by the UK {\it Swift} Science Data Centre at the University of Leicester. This work was supported by the National Natural Science Foundation of China (grants 10673034 and 10621303) and the National 973 Project on Fundamental Researches of China (2007CB815404 and 2009CB824800).

\begin{deluxetable}{lll|lll}
\tablecolumns{6}
\tablewidth{0pc}
\tablecaption{List of 150 {\it Swift} GRBs with Measured Redshift. \label{tab:sample}}
\tablehead{
\colhead{GRB} & \colhead{Redshift} & \colhead{$T_{90}$} & \colhead{GRB} & \colhead{Redshift}   & \colhead{$T_{90}$}}
\startdata
050126  & 1.29   & 48$\pm$22  & 050223  & 0.5915 & 23$\pm$5 \\
050315  & 1.949  & 96$\pm$14  & 050318  & 1.44   & 15$\pm$11 \\
050319  & 3.240  & 153$\pm$11 & 050401  & 2.90   & 33$\pm$1 \\
050416A & 0.6535 & 2.4$\pm$0.4    & 050505  & 4.27   & 59$\pm$7 \\
050509B & 0.226  & 0.05$\pm$0.02  & 050525A & 0.606  & 8.8$\pm$0.1 \\
050603  & 2.821  & 21$\pm$8   & 050724  & 0.258  & 96$\pm$8 \\
050730  & 3.967  & 157$\pm$18 & 050801  & 1.56   & 19$\pm$6 \\
050814  & 5.3    & 154$\pm$43 & 050820A & 2.612  & 245$\pm$23 \\
050824  & 0.83   & 23$\pm$4   & 050826  & 0.297  & 35$\pm$6  \\
050904  & 6.29   & 174$\pm$11 & 050908  & 3.344  & 17$\pm$4 \\
050922C & 2.198  & 4.5$\pm$0.5    & 051016B & 0.9364 & 4.0$\pm$0.5 \\
051109A & 2.346  & 37$\pm$7   & 051111  & 1.55   & 60$\pm$12 \\
051221A & 0.5465 & 1.4$\pm$0.2    & 060115  & 3.53   & 125$\pm$15 \\
060124  & 2.296  & 14$\pm$1   & 060202  & 0.783  & 199$\pm$23 \\
060206  & 4.048  & 7.6$\pm$1.9    &060210  & 3.91   & 255$\pm$122 \\
060218  & 0.0331 & \nodata          &060223A & 4.41   & 11$\pm$1 \\
060418  & 1.489  & 144$\pm$45 & 060502A & 1.51   & 28$\pm$10 \\
060510B & 4.9    & 271$\pm$21 &060512  & 2.1    & 8.5$\pm$1.6  \\
060522  & 5.11   & 64$\pm$6   & 060526  & 3.221  & 298$\pm$23 \\
060604  & 2.68   & 80$\pm$17        & 060605  & 3.78   & 115$\pm$35 \\
060607A & 3.082  & 99$\pm$31  &060614  & 0.125  & 106$\pm$3 \\
060707  & 3.425  & 61$\pm$6   & 060708  & 1.92   & 10$\pm$4  \\
060714  & 2.711  & 115$\pm$9  & 060729  & 0.54   & 115$\pm$34 \\
060801  & 1.30   & 0.5$\pm$0.1  & 060814  & 0.84   & 145$\pm$5 \\
060904B & 0.703  & 171$\pm$11 & 060906  & 3.686  & 44$\pm$8 \\
060908  & 1.8836 & 19$\pm$1   &060912A & 0.937  & 5.0$\pm$0.6 \\
060926  & 3.20   & 8.0$\pm$1.2    &060927  & 5.47   & 23$\pm$1 \\
061006  & 0.4377 & 130$\pm$30 & 061007  & 1.261  & 75$\pm$2  \\
061021  & 0.3463 & 46$\pm$5   & 061110A & 0.758  & 41$\pm$4 \\
061110B & 3.44   & 134$\pm$17 & 061121  & 1.314  & 81$\pm$46 \\
061126  & 1.1588 & 71$\pm$40  & 061210  & 0.4095 & 85$\pm$11 \\
061222A & 2.088  &  96$\pm$16       & 061222B & 3.355  & 40$\pm$6 \\
070110  & 2.352  & 88$\pm$14  & 070208  & 1.165  & 64$\pm$23  \\
070306  & 1.4959 & 209$\pm$65 & 070318  & 0.836  & 108$\pm$28 \\
070411  & 2.954  & 123$\pm$17 & 070419A & 0.97   & 160$\pm$51  \\
070506  & 2.31   & 4.3$\pm$0.5    & 070529  & 2.4996 & 89$\pm$20 \\
070611  & 2.04   & 12$\pm$3   & 070714B & 0.92   & 64$\pm$8 \\
070721B & 3.626  & 339$\pm$21 & 070724A & 0.457  & 0.7$\pm$0.2 \\
070802  & 2.45   & 17$\pm$3   & 070810A & 2.17   & 11$\pm$5 \\
071003  & 1.6043 & 148$\pm$4  & 071010A & 0.98   & 6.2$\pm$1.6 \\
071010B & 0.947  & 36$\pm$2   & 071020  & 2.145  & 4.2$\pm$0.6  \\
071025  & 5.2    & 238$\pm$36 & 071031  & 2.692  & 181$\pm$31 \\
071117  & 1.331  & 6.5$\pm$1.8    & 071122  & 1.14   & 69$\pm$14 \\
071227  & 0.383  & 303$\pm$202& 080129  & 4.349  & 48$\pm$36 \\
080210  & 2.641  & 45$\pm$11  & 080310  & 2.42   & 363$\pm$17 \\
080319B & 0.937  & 45.1$\pm$0.2   & 080319C & 1.95   & 34$\pm$9 \\
080330  & 1.51   & 61$\pm$6   & 080411  & 1.03   & 56$\pm$1 \\
080413A & 2.433  & 46$\pm$1   & 080430  & 0.767  & 14$\pm$2 \\
080520  & 1.545  & 2.8$\pm$0.7    & 080603B & 2.69   & 59$\pm$2 \\
080604  & 1.416  & 82$\pm$13  & 080605  & 1.6398 & 19$\pm$1 \\
080607  & 3.036  & 79$\pm$3   & 080707  & 1.23   & 32$\pm$6 \\
080710  & 0.845  & 120$\pm$17 & 080721  & 2.591  & 64$\pm$144 \\
080804  & 2.2045 & 34$\pm$16  & 080805  & 1.505  & 106$\pm$18  \\
080810  & 3.35   & 109$\pm$4  & 080905B & 2.374  & 94$\pm$10 \\
080913  & 6.695  & 7.9$\pm$0.9    & 080916A & 0.689  & 60$\pm$6 \\
080928  & 1.692  & 279$\pm$30 & 081007  & 0.5295 & 9.1$\pm$2.9 \\
081008  & 1.9685 & 186$\pm$40 & 081028  & 3.038  & 283$\pm$33 \\
081029  & 3.8479 & 270$\pm$46 & 081118  & 2.58   & 67$\pm$28 \\
081121  & 2.512  & 18$\pm$2   & 081203A & 2.05   & 221$\pm$83  \\
081222  & 2.77   & 25$\pm$3   & 090102  & 1.547  & 27$\pm$2 \\
090205  & 4.6497 & 8.8$\pm$1.8    & 090313  & 3.375  & 78$\pm$19 \\
090417B & 0.345  & 289$\pm$42 & 090418A & 1.608  & 56$\pm$4 \\
090423  & 8.26   & 10$\pm$1   & 090424  & 0.544  & 48$\pm$3 \\
090510  & 0.903  & 5.6$\pm$1.7    & 090516  & 4.109  & 208$\pm$64 \\
090519  & 3.85   & 64$\pm$11  & 090529  & 2.625  & 69$\pm$11 \\
090618  & 0.54   & 113$\pm$1  & 090715B & 3.0    & 266$\pm$12 \\
090726  & 2.71   & 59$\pm$15  & 090809  & 2.737  & 5.4$\pm$1.2 \\
090812  & 2.452  & 67$\pm$15  & 090814A & 0.696  & 80$\pm$8 \\
090926B & 1.24   & 112$\pm$1  & 090927  & 1.37   & 2.2$\pm$0.4  \\
091018  & 0.971  & 4.3$\pm$0.6    & 091020  & 1.71   & 39$\pm$5 \\
091024  & 1.092  & 99$\pm$14  & 091029  & 2.752  & 39$\pm$5 \\
091127  & 0.49   & 7.0$\pm$0.2    & 091208B & 1.063  & 14$\pm$3 \\
100219A & 4.6667 & 19$\pm$5   & 100302A & 4.813  & 18$\pm$2 \\
100316B & 1.180  & 3.9$\pm$0.5    & 100316D & 0.059  & \nodata  \\
100418A & 0.6235 & 7.0$\pm$1.0    & 100425A & 1.755  & 41$\pm$4 \\
\enddata
\end{deluxetable}

\end{document}